\documentclass[aps,prb,twocolumn,floatfix,citeautoscript,nofootinbib,superscriptaddress]{revtex4-2}
\usepackage{amsbsy}
\usepackage{latexsym,epsfig,graphicx}
\usepackage{dcolumn}
\usepackage{graphicx}
\usepackage{subfigure}
\usepackage{comment}
\usepackage{color}
\usepackage{xcolor}
\usepackage{soul}
\usepackage{bm}
\usepackage{mathrsfs}
\usepackage{amsfonts}
\usepackage{amsmath}
\usepackage{amssymb}
\usepackage{xspace}
\usepackage{epstopdf}
\usepackage{tabularx}
\usepackage{longtable}
\usepackage[colorlinks=true, letterpaper=true, pdfstartview=FitV, linkcolor=blue, citecolor=blue, urlcolor=blue]{hyperref}
\usepackage[normalem]{ulem}

\makeatletter
\newsavebox{\@brx}
\newcommand{\llangle}[1][]{\savebox{\@brx}{\(\m@th{#1\langle}\)}%
  \mathopen{\copy\@brx\kern-0.5\wd\@brx\usebox{\@brx}}}
\newcommand{\rrangle}[1][]{\savebox{\@brx}{\(\m@th{#1\rangle}\)}%
  \mathclose{\copy\@brx\kern-0.5\wd\@brx\usebox{\@brx}}}
\makeatother

\begin{document}
\title{Finite-temperature topological magnons in honeycomb ferromagnets with sublattice asymmetries}
\author{Lin-Gang Wei}
\affiliation{Department of Physics, College of Physical Science and Technology, Xiamen University, Xiamen 361005, China}
\author{Yun-Mei Li}
\email[contact author:]{yunmeili@xmu.edu.cn}
\affiliation{Department of Physics, College of Physical Science and Technology, Xiamen University, Xiamen 361005, China}

\begin{abstract}
  The \textit{Comment} [\href{10.1103/PhysRevLett.132.219601}{Phys. Rev. Lett. 132, 219601 (2024)}]
  pointed out that it is incorrect to predict the temperature-driven topological phase transition of Dirac magnons
  in honeycomb ferromagnets with Dzyaloshinskii-Moriya interactions based on the theory in \href{10.1103/PhysRevLett.127.217202}{Phys. Rev. Lett. 127, 217202 (2021)}. Here we propose that by breaking the sublattice symmetries in honeycomb ferromagnets, increasing temperature could induce topological transitions from trivial phase at zero temperature based on the linear spin wave theory to Chern insulating phase above a critical temperature without changing any spin-spin interactions. The key to the finite temperature topological magnons is considering the magnon-magnon interactions (MMIs) at a mean-field level. A self-consistently renormalized spin wave theory is employed to include self-energy corrections from MMIs, guaranteeing that the critical temperatures for topological transitions are below the Curi\'{e} temperatures. Across the critical temperatures, the magnon bandgap closes and reopens at $\mathbf{K}$ or $\mathbf{K}^{\prime}$ points in the Brillouin zone, accompanied by nontrivial Berry curvature transitions. However, in stark contrast to the work [\href{10.1103/PhysRevLett.127.217202}{Phys. Rev. Lett. 127, 217202 (2021)}], the topological transitions can not be revealed by the thermal Hall effect of magnons. Our work provides a realistic scheme for achieving finite-temperature topological phase in honeycomb ferromagnets.
\end{abstract}

\maketitle

\section{Introduction}

Last two decades have witnessed the remarkable progresses in the study of topological states of matter
~\cite{HasanMZ,QiXiaoLiang,HaldaneFDuncanM,Kosterlitz}.
Topological band theory is firstly proposed and widely studied in electronic systems
~\cite{HaldaneFDM,KaneCL1,KaneCL2,BAndreiBernevig,FuLiang,FuLiang2,RuiYu,CuiZuChang,MiaoMS,ZhangDong,SYXu,GarateIon}.
The integer topological invariants such as the Chern number indicate the nontrivial band structures and the existence of topologically protected edge states,
which has been proposed for applications in, for example, spintronics and quantum computations.
The definition of the topological invariants is independent on the temperature.
The thermal fluctuations at finite temperatures is usually thought to destabilize the topological phase
and break the topology-related transport properties, for example, the observations of quantized Hall conductance usually demand the low temperatures.

Recently, most efforts in this field have been devoted to extend topological concepts to photon~\cite{HaldaneFDM2,OzawaTomoki},
phonon~\cite{SebastiaDH}, polariton~\cite{KarzigTorsten}, and also the magnetic system.
Magnons, quantized spin excitations in magnets, have also been proposed to host nontrivial topological phases,
realized in magnets with artificially designed
structures~\cite{ShindouR1,ShindouR2,LiYM1,HuZhongqiang,KeShasha,LiYM2,KeShasha2}
or with special crystal symmetries~\cite{ZhangLifa,ChisnellR,KimSeKwon,KondoHiroki,LiYM3,CorticelliAlberto,BhowmickD,SunHao}. 
The emergence of edge or surface states immune to disorder and back scattering has great potential
for designing magnonic devices with low dissipation and power consumption.

Most theoretical studies on topological magnons rely on the celebrated linear spin wave theory (LSWT),
which provides a foundational framework but neglecting the impacts of MMIs.
The MMIs introduce temperature-dependent nonlinear corrections to the magnon spectrum
in the presence of the magnetic order~\cite{DysonFJ,OguchiT,LiuSH,BlochM,PDLoly1971,Lado2017,ZhengluLi,ERastelli1974,MGPini1981,WeiBin,MkhitaryanVV}.
In the year 2021, two works proposed that the MMIs can drive topological phase transitions of Dirac magnons between two nontrivial phases
with opposite Chern numbers at finite temperatures in honeycomb ferromagnets with Dzyaloshinskii-Moriya interactions(DMI)~\cite{LuYuShan,MAlexander}. 
The thermal Hall effect of magnons reveals the topological phase transitions by observing the sign change of the thermal Hall conductivity.
A subsequent work~\cite{LiYM4} predicted a universal scheme to achieve finite temperature topological phases of magnons
in two-dimensional antiferromagnets by considering the dipole-dipole interactions
and assuming a sublattice asymmetric next-nearest-neighboring exchange interactions.
In 2024, a \textit{Comment} paper~\cite{LiYM5} pointed out that the theory in Ref.~\cite{LuYuShan} is incorrect since
the critical temperatures for topological phase transitions coincide with the Curi\'{e} temperatures for magnetic phase transitions, above which
the magnon picture for the excitation spectrum is no longer valid. The method adopted in Ref.~\cite{LuYuShan} to deal with the MMIs is only 
valid when the temperature is far lower smaller than the Curi\'{e} temperature.

In order to discuss the renormalization of magnons at a wider temperature range and even near the Curi\'{e} temperature, 
self-consistent theory is usually applied~\cite{BlochM,PDLoly1971,Lado2017,ZhengluLi,ERastelli1974,MGPini1981,WeiBin,MkhitaryanVV}, which is 
an extension of the standard spin wave analysis of the excitations in magnets. 
The quartic operator terms are treated in a Hartree-Fock-like decoupling approximation, yielding quadratic corrections to Hamiltonian involving renormalization factors depending on the temperature. Self-consistent relations formed in this treatment since the truncated quadratic terms renormalize the magnon bands but depend on the renormalized magnon bands simultaneously.
The Hamiltonian and relevant physical quantities are solved self-consistently.

In this paper, we revisit the topology of magnons in honeycomb ferromagnets with DMI. In contrast to the previous works~\cite{LuYuShan,SunHao,BhowmickD,KimSeKwon,MAlexander}, 
we introduce sublattice symmetry breaking terms realized by introducing sublattice asymmetric easy-axis anisotropic term or the different sublattice magnetization, firstly proposed in a previous work~\cite{KimHongseok}. While this work studied the topological magnons within the framework of linear spin wave theory, we here include the impact of MMIs on the topological phases.

The self-consistently renormalized spin-wave theory (SRSWT) is employed to deal with the MMIs, guaranteeing the validity of our theory and credibility of the results.
This model fully manifests the magnonic analog of Haldane model~\cite{HaldaneFDM}. The topological invariant is the Chern number.
At zero temperature without considering the MMIs, we present the phase diagram and find nontrivial topological phase only exist in limited region of the parameter space.
While the finite temperature reveals the effects of MMIs, the phase boundaries between the trivial and nontrivial regions shifts, rendering
some certain zero-temperature trivial regions into nontrivial. This directly indicates that increasing temperature can induce topological transitions.
As a result of the temperature driven topological transitions, the gapless chiral edge states emerge above the critical temperatures.
The critical temperatures are lower than the Curi\'{e} temperatures. across the critical temperatures, the magnon bandgap closes and reopens at
K or K$^{\prime}$ points in the Brillouin zone. The Berry curvature experiences a sign change in the vicinity of these points.
By computing the thermal Hall conductivities with respect to the temperature, we do not find the obvious jump signatures across the critical temperatures.
Detecting topological phase transitions through the thermal Hall effect is not experimentally feasible for our model.

The organization of the remainder is as follows. In Sec.~\uppercase\expandafter{\romannumeral2}, we present the model and method utilizing the SRSWT to address the MMIs.
In Sec.~\uppercase\expandafter{\romannumeral3}, we present the phase diagram at both zero and finite temperatures. The results on the topological phase transitions,
the emergence of edge states and thermal Hall effect of magnons are also presented. Finally, in Sec.~\uppercase\expandafter{\romannumeral4}, we summarize our findings.

\section{Model and Methodology}

\begin{figure}[t]
  \centering
  \includegraphics[width=0.48\textwidth]{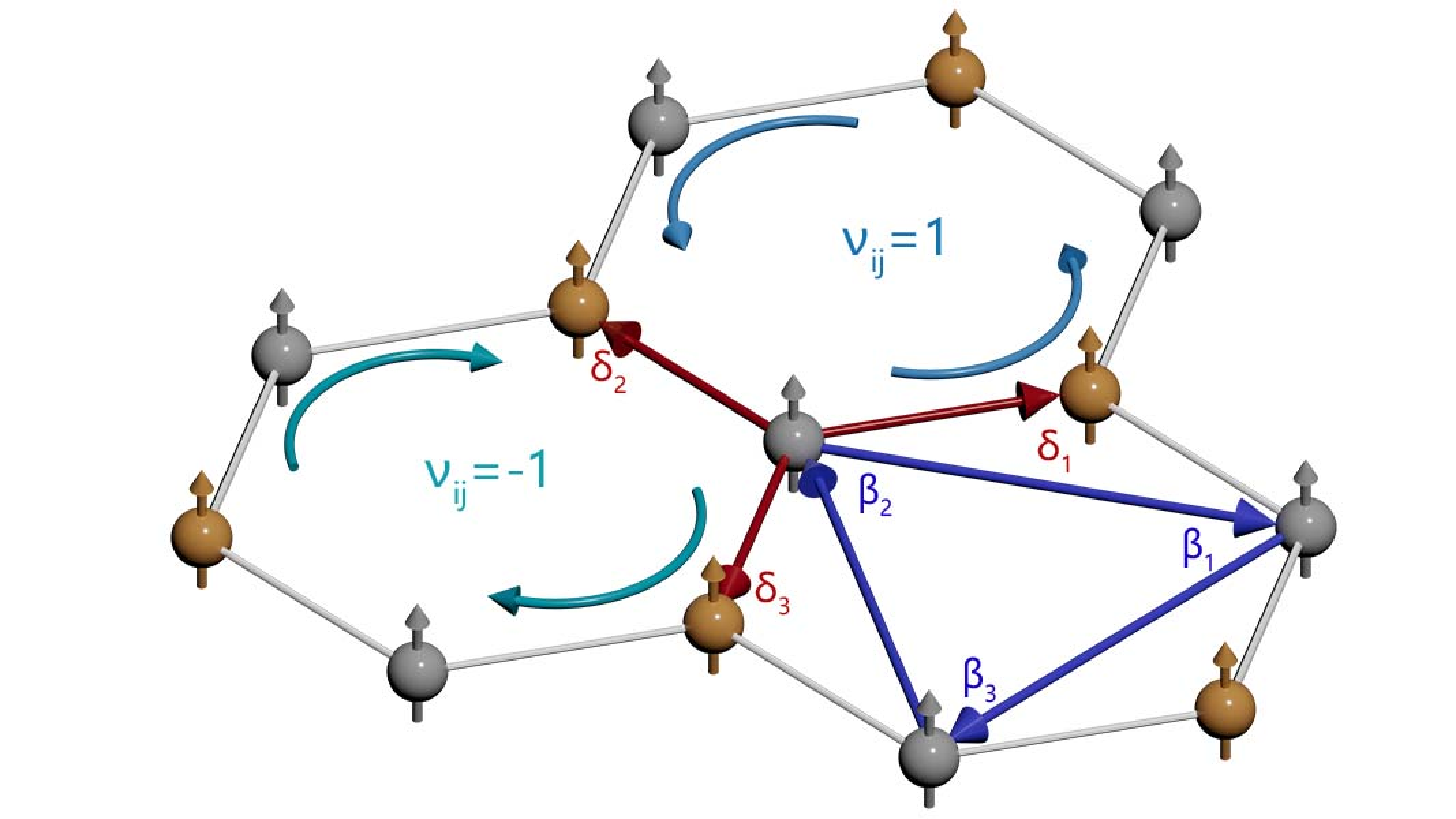}
  \caption{The honeycomb ferromagnets with NNN DMIs. The candidate materials includes the CrI$_{3}$, VI$_{3}$, etc.
  $\nu_{ij}=\pm 1$ characterizes the sign of DMIs with the sign dependent on the relative positions between two local spins, illustrated
  in the figure. $\delta_{i}$ and $\beta_{i}$ ($i$=1,2,3) denote the three nearest neighboring and next-nearest-neighboring vectors, respectively.}\label{fig1}
\end{figure}

We consider a honeycomb ferromagnet with second-nearest neighboring DMI and sublattice asymmetric easy-axis anistropic terms. The spin-spin interaction Hamiltonian is given by
\begin{eqnarray}
	H = &&J\sum_{\langle ij \rangle}\mathbf{S}_{i}\cdot\mathbf{S}_{j}
	+I\sum_{\langle ij \rangle}S_{i}^{z}S_{j}^{z} \nonumber\\
	&&+ D\sum_{\langle\langle ij \rangle\rangle}
	\nu_{ij}\mathbf{z}\cdot(\mathbf{S}_{i}\times\mathbf{S}_{j})- \sum_{i}K_i(S_{i}^z)^2.  \label{eq1}
\end{eqnarray}
The first term is the isotropic Heisenberg exchange interaction between nearest-neighbor spin with $J<0$,
while the second term denotes Ising-type anistropic exchange interaction with $I<0$.
The third term denotes the next-nearest-neighboring (NNN) DMI,
with $\nu_{ij}=\pm 1$ dependent on the relative position between NNN spins, as illustrated in Fig.~\ref{fig1}.
The last term is the easy-axis anisotropic term. We here assume different easy-axis anisotropy on the two sublattices,
which can be achieved by the substrate or in the homostructures and heterostructures~\cite{SivadasNikhil,HidalgoSacotoR,XiaoFeiping}.
We also do not restrict the sublattice magnetization to be the same. Therefore, the sublattice symmetry is broken.

We perform the Holstein-Primakoff (HP) transformations to get the magnon excitations and the MMIs.
We here expand the HP transformations up to the three bosonic operators
to consider the lowest-order MMIs,
\begin{align}\label{eq2}
 \begin{cases}
	S^{+}_{i} = S_{i}^{x}+iS_{i}^y\approx\sqrt{2S_{i}}(1-\frac{ a_{i}^{\dagger}a_{i} }{4S_{i}})a_{i}\\
	S^{-}_{i}=S_{i}^{x}-iS_{i}^y\approx\sqrt{2S_{i}}a_{i}^{\dagger}(1-\frac{a_{i}^{\dagger}a_{i}}{4S_{i}})\\
	S^{z}_{i}=S_{i}-a_{i}^{\dagger}a_{i}
\end{cases},
\end{align}
$a_{i}(a_{i}^\dagger)$ is the magnon annihilation (creation) operator and obeys commutation relation $[a_i,a_j^\dagger]=\delta_{i j}$.
The Hamiltonian in Eq.~\eqref{eq1} can be expanded as $H=E_{0}+H_{2}+H_{4}$. $E_{0}$ is the ground state energy.
$H_{2}$ and $H_{4}$ denote the terms with two and four magnon operators, respectively. $H_{2}$ is the magnon Hamiltonian from linear spin wave theory, while $H_{4}$ describes the lowest-order magnon-magnon interactions.
We do not present the complicated expressions of $H_{2}$ and $H_{4}$ in real space here
and we discuss them in the momentum space subsequently instead.

To discuss the magnon dispersions and the effect of MMIs in the momentum space,
one should perform the Fourier transforms for a system
of $N$ unit cells under periodic boundary condition,
\begin{equation}\label{eq3}
    a_{i}^\dagger =\frac{1}{\sqrt{N}}\sum_{\mathbf{k}} e^{-i\mathbf{k} \cdot \mathbf{r_i}} b_{\mu\mathbf{k}}^\dagger,
\end{equation}
$\mu=1,2$ mark the A and B sublattice index. Using the relation
$\sum_{\mathbf{r}_{i}}e^{i(\mathbf{k}-\mathbf{k}^{\prime})\cdot\mathbf{r}_{i}}=N\delta_{\mathbf{k}\mathbf{k}^{\prime}}$,
$H_{2}$ and $H_{4}$ can be transformed into the $\mathbf{k}$-space.

The two operators term can be expressed as $H_{2}=\sum_{\mathbf{k}}\Psi^{\dagger}_{\mathbf{k}}H_{2\mathbf{k}}\Psi_{\mathbf{k}}$,
\begin{equation} \label{eq4}
    H_{2\mathbf{k}}=h_{0}\mathbb{I}_{2}+\mathbf{h}\cdot\bm\sigma,
\end{equation}
where $\Psi_{\mathbf{k}}=(b_{1\mathbf{k}},b_{2\mathbf{k}})^{T}$ and $\bm\sigma$ are the vectors of Pauli matrices acting on the sublattice index.
$h_{0}=-3(J+I)S_{0}+K_{1}(S_{1}-\frac{1}{2})+K_{2}(S_{2}-\frac{1}{2})+\frac{1}{2}D g(\mathbf{k})\Delta S$,
$h_x =-JS\text{Re}(f({\mathbf{k}})$, $h_y =-JS\text{Im}(f_{\mathbf{k}})$,
$h_z =-\frac{3}{2}(J+I)\Delta S+K_{1}(S_{1}-\frac{1}{2})-K_{2}(S_{2}-\frac{1}{2}) +D S_{0} g(\mathbf{k})$,
$f(\mathbf{k})=\sum_{\mathbf{\delta}_{i}} e^{i\mathbf{k}\cdot \mathbf{\delta}_{i}}$, $g(\mathbf{k})=2\sum_{\mathbf{\beta_{i}}}\sin(\mathbf{k}\cdot\mathbf{\beta}_{i})$, $\mathbf{\delta}_i$ and $\mathbf{\beta}_i$ are three nearest-neighbor and next-nearest-neighboring vectors, illustrated in Fig.~\ref{fig1}.
$S_{0}=(S_{1}+S_{2})/2$, $S=\sqrt{S_{1}S_{2}}$, $\Delta S=S_{1}-S_{2}$.

We now discuss the effect of MMIs. The Hamiltonian describing MMIs in the momentum space reads
\begin{widetext}
\begin{eqnarray}
    H_{4}=&&-\frac{1}{4N}\sum_{\{\mathbf{k}_i\}}\left\{
    \sum_{\mu}\left[4K_{\mu} + \frac{g(\mathbf{k}_1) + g(\mathbf{k}_4)}{2}D\right]b_{\mu \mathbf{k}_1}^\dagger  b_{\mu \mathbf{k}_2}^\dagger  b_{\mu \mathbf{k}_3}  b_{\mu \mathbf{k}_4}
    +\sum_{\mu\neq \nu} \left[\frac{J+I}{2}h_{12}(\mathbf{k}_4-\mathbf{k}_2) b_{\mu \mathbf{k}_1}^\dagger  b_{\nu \mathbf{k}_2}^\dagger  b_{\mu \mathbf{k}_3}  b_{\nu \mathbf{k}_4}\right.\right.     \nonumber\\
    &&\left. \left.-\frac{JS}{S_\mu}h_{\mu \nu}(\mathbf{k}_4)b_{\mu \mathbf{k}_1}^\dagger  b_{\mu \mathbf{k}_2}^\dagger  b_{\mu \mathbf{k}_3}  b_{\nu \mathbf{k}_4} - \frac{JS}{S_\nu}h_{\mu \nu}(\mathbf{k}_1)
    b_{\mu \mathbf{k}_1}^\dagger  b_{\nu \mathbf{k}_{2}}^\dagger  b_{\nu \mathbf{k}_{3}}  b_{\nu \mathbf{k}_4}\right] \right\}\delta_{\mathbf{k}_1+\mathbf{k}_{2} , \mathbf{k}_{3}+\mathbf{k}_{4}}, \label{eq5}
\end{eqnarray}
\end{widetext}
where $h_{\mu \nu}(\mathbf{k})=\sum_{\mathbf{\delta}_i}e^{i(\nu - \mu)\mathbf{k}\cdot\mathbf{\delta}_i}$.
To study the effect of the MMIs on the magnon spectrum, we apply the SRSWT to deal with the MMIs
through Hartree-Fock-like decoupling,
\begin{eqnarray}\label{eq6}
	&&b_{\mu k_1}^\dagger  b_{\mu' k_2}^\dagger  b_{\nu k_3}  b_{\nu' k_4} \nonumber\\
	&\approx&
	\langle   b_{\mu k_1}^\dagger  b_{\nu k_3} \rangle   b_{\mu' k_2}^\dagger  b_{\nu' k_4}
	+\langle   b_{\mu k_1}^\dagger  b_{\nu' k_4} \rangle   b_{\mu' k_2}^\dagger  b_{\nu k_3}\nonumber\\
	&+&\langle   b_{\mu' k_2}^\dagger  b_{\nu k_3} \rangle   b_{\mu k_1}^\dagger  b_{\nu' k_4}
	+\langle   b_{\mu' k_2}^\dagger  b_{\nu' k_4} \rangle   b_{\mu k_1}^\dagger  b_{\nu k_3}.
\end{eqnarray}
Furthermore, we keep only the diagonal terms,
\begin{equation}\label{eq7}
    \langle   b_{\mu k}^\dagger  b_{\mu' k'} \rangle = \delta_{kk'}\langle   b_{\mu k}^\dagger  b_{\mu' k} \rangle.
\end{equation}
By applying the above mean-field approximation, we get the effective spin-wave Hamiltonian $H^{\mathrm{eff}}=H_{2}+H_{4}^{\mathrm{eff}}$, and  $H^{\mathrm{eff}}=\sum_{\mathbf{k}}\Psi^{\dagger}_{\mathbf{k}}H_{\mathbf{k}}^{\mathrm{eff}}\Psi_{\mathbf{k}}$
\begin{eqnarray}\label{eq8}
	H_{\mathbf{k}}^{\mathrm{eff}}=\begin{pmatrix}
		A_1(\mathbf{k}) & B_1(\mathbf{k})\\
		B_2(\mathbf{k}) & A_2(\mathbf{k})
	\end{pmatrix}
\end{eqnarray}
where
\begin{eqnarray}\label{eq9}
	A_1(\mathbf{k})=&&-[3 (J+I)\bar{S}_{22} + Jf_{1}]+D(\bar{S}_{11}g(\mathbf{k}) - g_{1})  \nonumber\\
		                         &&+(4\bar{S}_{11}-2S_1-1)K_1,    \nonumber\\
	A_2(\mathbf{k})=&&-[3 (J+I)\bar{S}_{11} + Jf_{2}]-D(\bar{S}_{22}g(\mathbf{k}) - g_{2}) \nonumber\\
		                         &&+(4\bar{S}_{22}-2S_2-1)K_2,      \nonumber\\
	B_1(\mathbf{k})=&&[J\bar{S}_{12} + (J+I)f_{12}]h_{12}(\mathbf{k}),    \nonumber\\
	B_2(\mathbf{k})=&&[J\bar{S}_{12} + (J+I)f_{21}]h_{21}(\mathbf{k}).
\end{eqnarray}
$f_{12}$, $f_{\mu}$, $g_{\mu}$ and $\bar{S}_{\mu \nu}$ in~\eqref{eq9} are given by
\begin{eqnarray}\label{eq10}
    \bar{S}_{\mu \nu} &=& \sqrt{S_{\mu}S_\nu}\left[1 - \frac{1}{2N}\sum_{\mathbf{k}}\left(\frac{\langle   b_{\mu\textbf{k}}^\dagger  b_{\mu\textbf{k}} \rangle }{S_\mu}+ \frac{\langle   b_{\nu\textbf{k}}^\dagger  b_{\nu\textbf{k}}\rangle}{S_\nu} \right)\right], \nonumber\\
	f_{\mu} &=& \frac{S}{NS_{\mu}}\textrm{Re}\left[h_{12}(\mathbf{k})\langle   b_{1k}^\dagger  b_{2k} \rangle \right],     \nonumber\\
	f_{12} &=& f_{21}^{*} = \frac{1}{3N}\sum_{\mathbf{k}}f(\mathbf{k})\langle   b_{2k}^\dagger  b_{1k} \rangle,   \nonumber\\
	g_{\mu} &=& \sum_{\textbf{k}} g(\mathbf{k} )\langle   b_{\mu \textbf{k}}^\dagger  b_{\mu \textbf{k}} \rangle.
\end{eqnarray}
$\langle b_{\mu \mathbf{k}}^\dagger b_{\nu \mathbf{k}} \rangle$ means the thermal average of the two-magnon operators.
The sublattice magnetization $\bar{S}_{1}=\bar{S}_{11}$, $\bar{S}_{2}=\bar{S}_{22}$.
The self-consistent equations are built by expressing the
above thermodynamical quantities through the Hamiltonian
defined in Eq.~\eqref{eq8}.  We define the creation-annihilation operators
$\alpha_{\mathbf{k}n}^{\dagger}$, $\alpha_{\mathbf{k}n}$ of magnon eigenmodes
of Hamiltonian $H_{\mathbf{k}}^{\mathrm{eff}}$, where $n=1,2$ label the magnon branches.
$\alpha_{\mathbf{k}n}^{\dagger}$, and $\alpha_{\mathbf{k}n}$ are linear combinations of $b_{\mu\mathbf{k}}^{\dagger}$,
$b_{\mu\mathbf{k}}$. We have
\begin{equation}\label{eq11}
  b_{\mu\mathbf{k}}=\sum_{n=1}^{2}[\Lambda_{\mathbf{k}}]_{\mu n}\alpha_{\mathbf{k}n},
\end{equation}
and the corresponding complex conjugation relation between $\alpha_{\mathbf{k}n}^{\dagger}$ and $b_{\mu\mathbf{k}}^{\dagger}$.
$\Lambda_{\mathbf{k}}$ is the eigenvector matrix diagonalizing $H_{\mathbf{k}}^{\mathrm{eff}}$,
$\Lambda_{\mathbf{k}}^{\dagger}H_{\mathbf{k}}^{\mathrm{eff}}\Lambda_{\mathbf{k}}=\textrm{diag}\{E_{\mathbf{kn}}\}$, where
$E_{\mathbf{kn}}$ is the eigenvalues and $\Lambda_{\mathbf{k}}^{\dagger}\Lambda_{\mathbf{k}}=\mathbb{I}_{2}$.
With the relation in Eq.~\eqref{eq11} and
$\rho_{\mathbf{k}n}=\langle\alpha_{\mathbf{k}n}^{\dagger}\alpha_{\mathbf{k}n} \rangle=[\exp(\beta E_{\mathbf{k}n})-1]^{-1}$,
one have
\begin{equation}\label{eq12}
  \langle b_{\mu \mathbf{k}}^\dagger b_{\nu \mathbf{k}}\rangle=\sum_{n=1}^{2}
  [\Lambda_\mathbf{k}^*]_{\mu n}[\Lambda_{\mathbf{k}}]_{\nu n}\rho_{\mathbf{k}n}.
\end{equation}
$\beta=1/k_{B}T$ and $T$ is the temperature, $k_{B}$ is the Boltzmann constant.
The Eqs.~\eqref{eq8},  \eqref{eq9}, \eqref{eq10}, \eqref{eq11}, \eqref{eq12} form the self-consistent relations. 
We can numerically calculate the thermodynamical quantities and get the effective Hamiltonian for magnons at given finite temperatures.
We perform the calculations and discuss the topological properties of magnon in the temperature regime giving converged results after sufficiently large 
iterations.  

The effective Hamiltonian indicates a magnonic analog of the Haldane model on the Chern insulators.
The topological invariant is the Chern number, defined as
\begin{equation}\label{eq13}
	C_n=\frac{1}{2\pi}\int_{BZ}\Omega_{\mathbf{k}n}^{z}d^2\mathbf{k},
\end{equation}
where
\begin{eqnarray}\label{eq14}
  \Omega_{\mathbf{k}n}^{z}=-2\textrm{Im}\sum_{\substack{\mu\mu^{\prime}\\ \nu\nu^{\prime}}}&&
  \frac{ [\Lambda_\mathbf{k}^*]_{\mu n}[\Lambda_{\mathbf{k}}]_{\nu m} [\Lambda_\mathbf{k}^*]_{\mu^{\prime} m}[\Lambda_{\mathbf{k}}]_{\nu^{\prime} n}}
  {(E_m-E_n)^2} \nonumber \\
  &\times& \frac{\partial H_{\mathbf{k},\mu\nu}^{\textrm{eff}}}{\partial k_{x}}\frac{\partial H_{\mathbf{k},\mu^{\prime}\nu^{\prime}}^{\textrm{eff}}}{\partial k_{y}}, \quad m \neq n
\end{eqnarray}
is the Berry curvature of $n$-th band~\cite{XiaoDi}.
Previous works proposed that the topological phase transition of magnon can be revealed
through the thermal Hall effect due to the nontrivial Berry curvature distribution in the momentum space~\cite{LuYuShan,MAlexander,LiYM4}.
The thermal Hall conductivity of magnons is given by~\cite{MatsumotoRyo}
\begin{equation}\label{eq15}
	\kappa_{xy}=-\frac{k_{B}^{2}T}{(2\pi)^2\hbar}\sum_{n}\int d^2\mathbf{k}\Omega_{\mathbf{k}n}^{z}c_2(\rho_{\mathbf{k}n}),
\end{equation}
where $c_2(x) = (1+x)(\ln\frac{1+x}{x})^2-(\ln x)^2-2\text{Li}_2(-x)$, and $\text{Li}_2(x)$ the polylogarithm function.

\section{Results and Discussions}

\subsection{The case $K_{1}\neq K_{2}$ and $S_{1}=S_{2}=S$ }
\begin{figure}[t]
  \centering
  \includegraphics[width=0.5\textwidth]{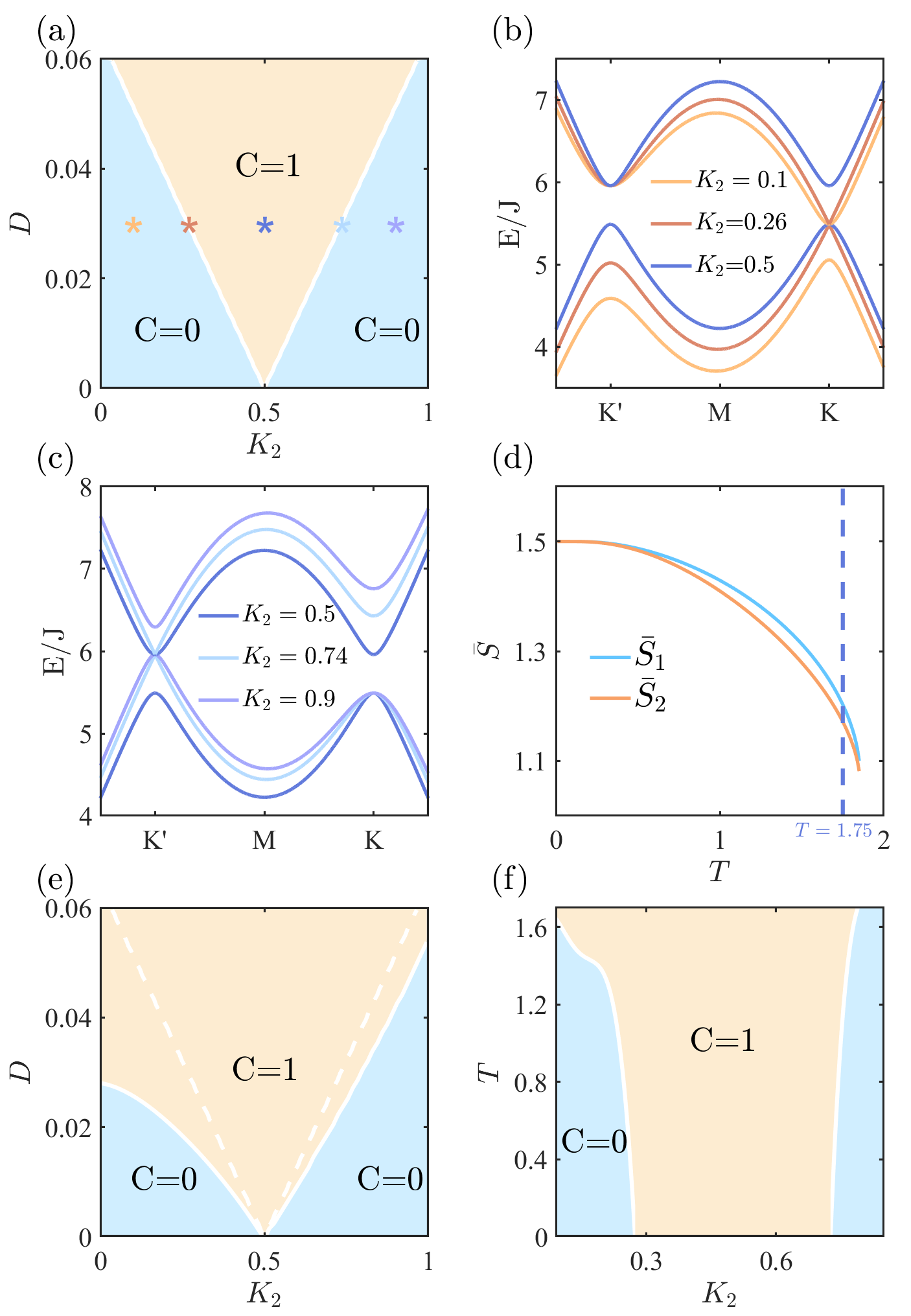}
  \caption{(a) The phase diagram in the $K_{2}-D$ plane at zero temperature $T=0$. The order parameter is the Chern number of the upper magnon band.
   (b) The magnon bands at  $T=0$ with three different value of $K_{2}$. The bandgap closes and reopens near $\mathbf{K}$ point.
   (c) The magnon bands at $T=0$ with another three larger $K_{2}$. The bandgap closes and reopens near $\mathbf{K}^{\prime}$ point.
   In (b) and (c), $D=0.03$.
   (d) The sublattice magnetization with respect to the temperature for $K_{2}=0$.
   (e) The phase diagram in the $K_{2}-D$ plane at finite temperature $T=1.75$. The while dashed lines denote the phase boundaries at zero temperature from (a).
   (f) The phase diagram in the $K_2-T$ plane at given DMI strength $D=0.03$.
   In all the panels, the other parameters are given by $S_{1} = S_{2} = 1.5$, $K_{1} = 0.5$, $I=-0.05$ and $J = -1$.} \label{fig2}
\end{figure}

We firstly present the results for $S_{1}=S_{2}$ with $K_{1}\neq K_{2}$.
At zero temperature, we do not need consider the effect of MMIs because the self-energy corrections vanish.
We can apply the Hamiltonian in Eq.~\eqref{eq4} based on the linear spin wave theory to discuss the topological properties of magnon bands.
In the absence of the sublattice asymmetric and DMI, the magnon bands are gapless, exhibiting two Dirac points at
$\mathbf{K}$ and $\mathbf{K}^{\prime}$. Both sublattice asymmetric easy-axis anisotropy and DMI can open the bandgap
at $\mathbf{K}$ and $\mathbf{K}^{\prime}$ points but in different schemes. The sublattice asymmetric easy-axis anisotropy $\Delta K=K_{1}-K_{2}\neq 0$,
breaks the inversion symmetry while the DMI breaks the time-reversal symmetry in Hamiltonian of Eq.~\eqref{eq4}.
$\Delta K$ gives the gap term with the same value at $\mathbf{K}$ and $\mathbf{K}^{\prime}$ in the Brillouin zone while the DMI gives opposite term.
Therefore the values of the bandgap at $\mathbf{K}$ and $\mathbf{K}^{\prime}$ are no longer the same any more.
Varying $\Delta K=K_{1}-K_{2}$ and $D$ could even close and reopen the bandgap at $\mathbf{K}$ or $\mathbf{K}^{\prime}$.
Using the relations $f(\mathbf{K})=f(\mathbf{K}^{\prime})=0$,
$g(\mathbf{K})=-g(\mathbf{K}^{\prime})=\frac{3\sqrt{3}}{2}$,
the gap closing condition is $h_{z}=0$, giving rise to
\begin{eqnarray}\label{eq16}
  \Delta K(S-\frac{1}{2}) + \frac{3\sqrt{3}}{2}DS  &=& 0     \quad  at \quad \mathbf{K}, \nonumber \\
  \Delta K(S-\frac{1}{2}) - \frac{3\sqrt{3}}{2}DS  &=& 0   \quad  at  \quad  \mathbf{K}^{\prime}.
\end{eqnarray}
Then we can get the analytical conditions, $ \Delta K(2S-1)=\pm 3\sqrt{3}DS$.
In Fig.~\ref{fig2} (a), we fix the value of $K_{1}$ and the phase boundary condition divides the $K_{2}-D$ plane into three parts.
In the region $K_{1}-\frac{3\sqrt{3}DS}{2S-1}<K_{2}<K_{1}+\frac{3\sqrt{3}DS}{2S-1}$, we find the Chern number
of the upper (lower) magnon band is $1$ (-1). Outside the region, the Chern number of both bands are $0$.
As shown in Fig.~\ref{fig2} (b), increasing $K_{2}$ from $0$, the bandgap at $\mathbf{K}$ point decreases firstly, closes at
$K_{1}-\frac{3\sqrt{3}DS}{2S-1}=K_{2}$ and then increases.
Further increasing $\Delta K$ closes the reopens the gap at $\mathbf{K}^{\prime}$ point, as shown in Fig.~\ref{fig2} (c).
These gap closings and reopenings result in the topological phase transitions between the $C=0$ and $C=\pm 1$ phases.

Increasing temperature is expected to make the above results based on the linear spin wave theory no longer valid since we should consider the self-energy corrections from the MMIs. The self-energy corrections $H_{4\mathbf{k}}^{\mathrm{eff}}$ is temperature dependent. We apply the self-consistently renormalized spin wave theory to consider the self-energy corrections.
The self-consistent calculations are adopted to determine the values of the quantities in Eq.~\eqref{eq10} at any given temperatures. The self-consistent calculations also help us to confirm
the temperature we choose is below the Curi\'{e} temperature. As shown in Fig.~\ref{fig2} (d), we present the sublattice magnetization evolution with respect to the temperature at $K_{2}=0$.
The Curi\'{e} temperature is about $T_{Curie}\simeq 1.8J$. The larger $K_{2}$ values would give larger Curi\'{e} temperatures.
Interestingly, the sublattice magnetization at finite temperature hold different values, arising from the sublattice asymmetric easy-axis anisotropy.

\begin{figure}[t]
  \centering
  \includegraphics[width=0.48\textwidth]{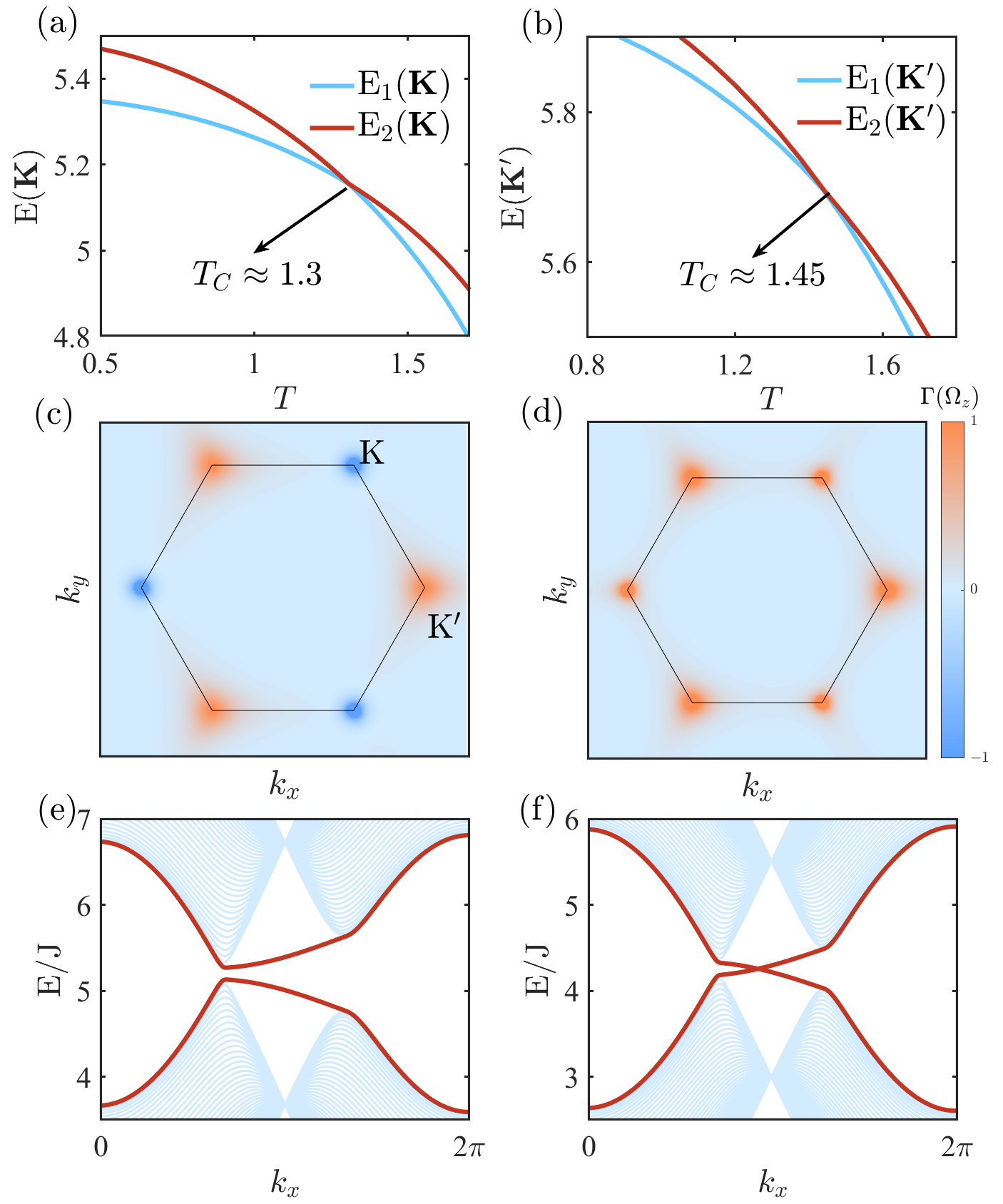}
  \caption{(a) The magnon energies at $\mathbf{K}$ point with respect to the temperature at $K_{2}=0.2$.
  (b) The magnon energies at $\mathbf{K}^{\prime}$ point with respect to the temperature at $K_{2}=0.7$.
  (c-d) The magnon Berry curvature distribution in log scale $\Gamma(\Omega_1^z) = \text{sign}(\Omega_1^z)\text{log}_{10}(1+|\Omega_1^z|)$
   at $T=0$ (c) and $T=1.75$ for $K_{2}=0.2$.
  (e-f) The zigzag ribbon band at $T=0$ and $T=1.75$ for $K_{2}=0.2$.}\label{fig3}
\end{figure} 

Same to the zero temperature case, the Chern number of the upper band is adopted as the order parameter.
The phase diagram in the $K_{2}-D$ plane at $T=1.75J$ is shown in Fig.~\ref{fig2} (e). Compared to the phase diagram at zero temperature, the phase boundaries between the
trivial and nontrivial regions are different. The phase boundaries shift to make the nontrivial region larger.
This directly indicates that there exist topological phase transitions from the trivial to nontrivial phases by increasing temperatures for some certain value of $K_{2}$.
To study the temperature induced topological transitions further, we plot the phase diagram in $K_{2}-T$ plane at a given $D$ in Fig.~\ref{fig2} (f).
$K_{2}<K_{1}-\frac{3\sqrt{3}DS}{2S-1}$ or  $K_{2}>K_{1}+\frac{3\sqrt{3}DS}{2S-1}$ give trivial phase at zero temperature. However,
increasing temperature will transform these two regions into nontrivial phase across critical temperatures.

The magnon energies at $\mathbf{K}$ point for a certain $K_{2}$ value smaller than $K_{1}-\frac{3\sqrt{3}DS}{2S-1}$ are plotted in Fig.~\ref{fig3} (a).
Across the critical temperature, the magnon gap closes and reopens. For $K_{2}>K_{1}+\frac{3\sqrt{3}DS}{2S-1}$,
the magnon gap closes and reopens at $\mathbf{K}^{\prime}$ point across the critical temperature [Fig.~\ref{fig3} (b)].
Accompanied by the gap closing the reopening, the Berry curvature experiences a sign change
near $\mathbf{K}$ ($\mathbf{K}^{\prime}$) point for $K_{2}<K_{1}-\frac{3\sqrt{3}DS}{2S-1}$ ($K_{2}>K_{1}+\frac{3\sqrt{3}DS}{2S-1}$), as
shown in Fig.~\ref{fig3} (c) and \ref{fig3} (d), making the Chern number jumps from $0$ to $\pm 1$.

\begin{figure}[t]
  \centering
  \includegraphics[width=0.5\textwidth]{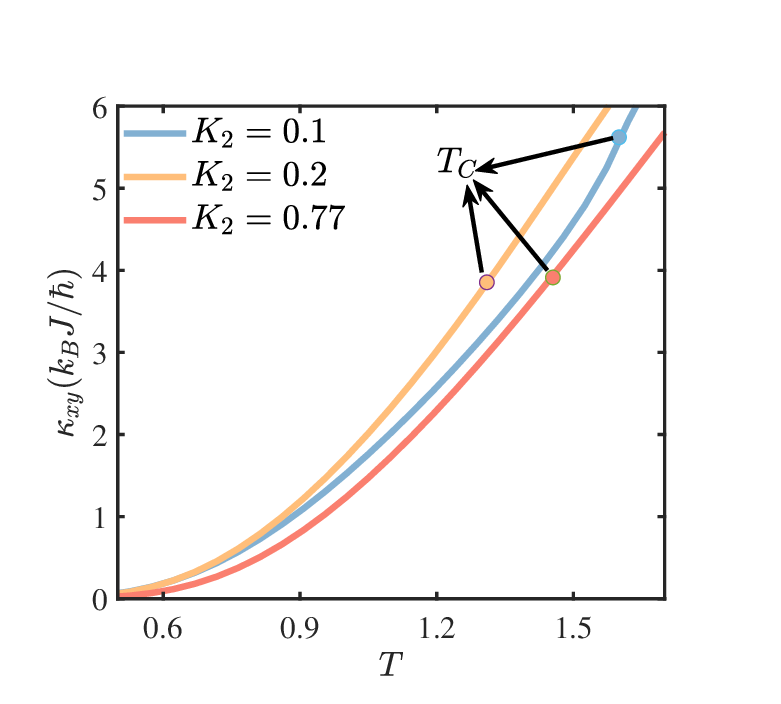}
  \caption{The thermal Hall conductivity in the unit $k_{B}J/\hbar$ with respect to the temperature for three different $K_{2}$.}\label{fig4}
\end{figure}

We give an explanation on the emergence of the topological transition at finite temperatures.
The two Hamiltonians in Eq.~\eqref{eq4} and \eqref{eq8} shares the same formalism with parameters are changed.
From Eq.~\eqref{eq8}, the gap closing condition at $\mathbf{K}$ is given by
\begin{eqnarray}
   &&3\sqrt{3}D\bar{S}_{0}+4(K_{1}\bar{S}_{1}-K_{2}\bar{S}_{2}) =  \nonumber \\
    && \Delta K(2S+1)+3(J+I)\Delta\bar{S}
\end{eqnarray}
while at $\mathbf{K}^{\prime}$ point is give by
\begin{eqnarray}
&& -3\sqrt{3}D\bar{S}_{0}+4(K_{1}\bar{S}_{1}-K_{2}\bar{S}_{2}) = \nonumber \\
&& \Delta K(2S+1)+3(J+I)\Delta\bar{S}
\end{eqnarray}
where $\bar{S}_{0}=(\bar{S}_{1}+\bar{S}_{2})/2$, $\Delta\bar{S}=\bar{S}_{1}-\bar{S}_{2}$.  Compared to Eq.~\eqref{eq16},
the complex gap closing conditions can only be satisfied at finite temperatures, giving rise to the topological magnons defined only above the critical temperatures.

The finite temperature topological phases result in the emergence of the gapless chiral edge states above critical temperatures. Fig.~\ref{fig3} (e) and \ref{fig3} (f) plot
the zigzag ribbon band structures at two temperature without changing any other parameters. One temperature is below  $T_{c}$.
The edge states arise from sublattice polarization in the zigzag ribbon. The other temperature is above $T_{c}$, presenting
the gapless chiral edge states. The emergence of gapless edge state can be used for characterizing the topological transitions.
The theory in Refs.~\cite{LuYuShan,MAlexander,LiYM4} predicted that the thermal Hall effect of magnons
due to the Berry curvature could be the signatures of the topological phase transitions.
The thermal Hall conductivities with respect to the temperature for three different $K_{2}$ values are computed and presented in Fig.~\ref{fig4}.
No clear signatures revealing the topological transitions are observed.

\subsection{The case $K_{1}\neq K_{2}$ and $S_{1}\neq S_{2}$}

\begin{figure}[!t]
  \centering
  \includegraphics[width=0.5\textwidth]{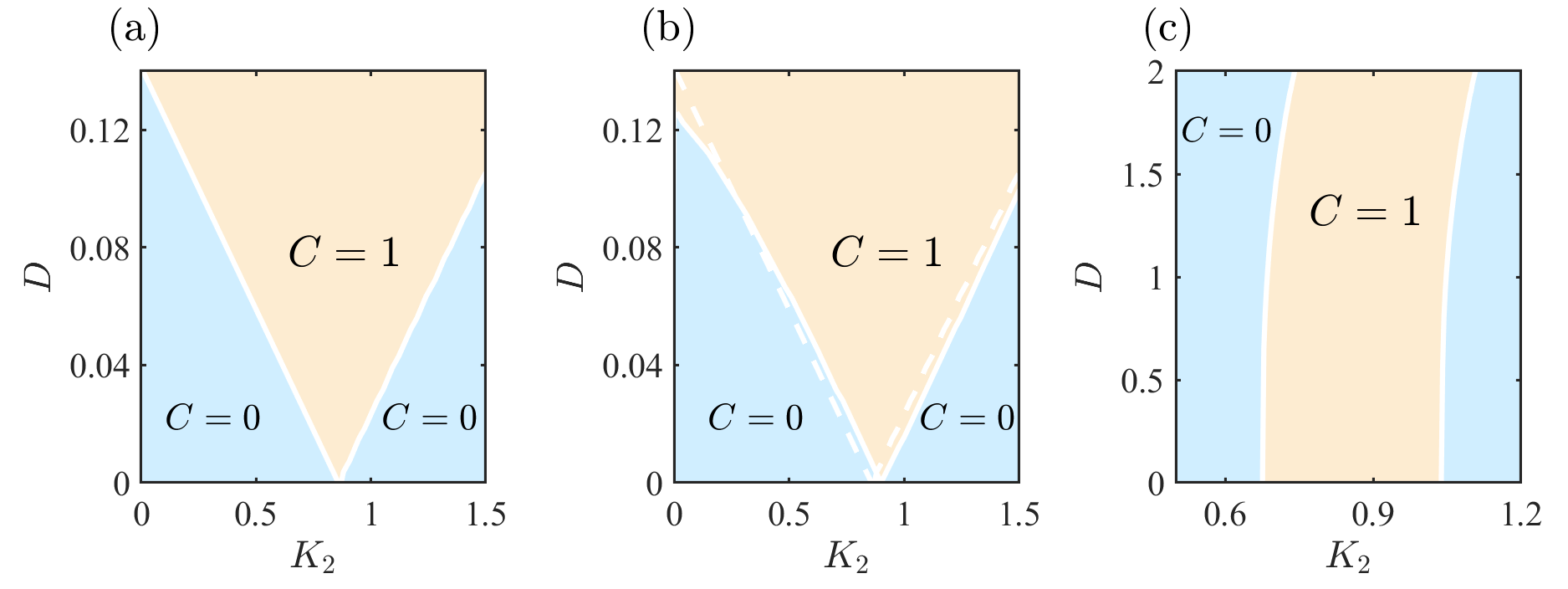}
  \caption{(a) The phase diagram in the $K_{2}-D$ plane at $T=0$.
  (b) The phase diagram in the $K_{2}-D$ plane at $T=1.75$ .
  (c) The phase diagram in the $K_2-T$ plane with $D=0.03$.
  We set $S_{1} =1.5$, $ S_{2}=2$ and $K_1 = 0.5$, $I=-0.05$, $J = -1$ for all calculations.}\label{fig5}
\end{figure}

We now consider the case for different sublattice magnetization, $S_{1}\neq S_{2}$. Same to the case
$S_{1}=S_{2}$, the phase boundaries between the trivial and nontrivial phases are determined by the gap closing condition.
At Zero temperature, the gap closing condition at $ \mathbf{K}$ point is given by
\begin{eqnarray}\label{eq17}
  \Delta K(2S_{0}-1)&+&(K _{1}+K_{2}-3J-3I)\Delta S   \nonumber  \\
  &+& 3\sqrt{3}DS_{0} = 0,
\end{eqnarray}
while at $\mathbf{K}^{\prime}$ point is given by 
\begin{eqnarray}\label{eq17}
  \Delta K(2S_{0}-1)&+&(K _{1}+K_{2}-3J-3I)\Delta S        \nonumber  \\
  &-& 3\sqrt{3}DS_{0} = 0. 
\end{eqnarray}
The phase diagram at zero temperature is shown in Fig.~\ref{fig5} (a).
When $K_{1}=K_{2}=K$, $D>D_{c}=\frac{[3(J+I)-2K]|\Delta S|}{3\sqrt{3}S_{0}}$ gives nontrivial phase.

Similar to the zero temperature, at finite temperatures, we can also get the phase boundaries between the trivial and nontrivial phases
by the conditions $A_{1}(\mathbf{K})=A_{2}(\mathbf{K})$ and  $A_{1}(\mathbf{K}^{\prime})=A_{2}(\mathbf{K}^{\prime})$.
The phase diagram at a given finite temperature is plotted  in  Fig.~\ref{fig5} (b). Comparing to  Fig.~\ref{fig5} (a), the phase boundaries shift.
The self-energy corrections from the MMIs also changes the topological properties of magnons at finite temperatures.
From the phase diagram in the $K_{2}-T$ plane [Fig.~\ref{fig5} (c)], increasing temperature not only  induces topological phase transitions from
trivial to nontrivial phase, but also make the nontrivial phase to be trivial across critical temperatures.
Increasing temperature could suppress the topological phase.
These topological phase transitions are all due to the bandgap closing and reopening at $\mathbf{K}$ or $\mathbf{K}^{\prime}$ points in the Brillouin zone, accompanied by the Berry curvature transition, the emergence or destruction of the chiral edge states, similar to the case $S_{1}=S_{2}$. 
Besides, we did not find the realistic materials corresponding to ferromagnets with different sublattice magnetizations.
Therefore, we here made brief discussions on the case $S_{1}\neq S_{2}$.
 
\subsection{Higher-order vertex corrections}
\begin{figure}[t]
	\centering
	\includegraphics[width=0.48\textwidth]{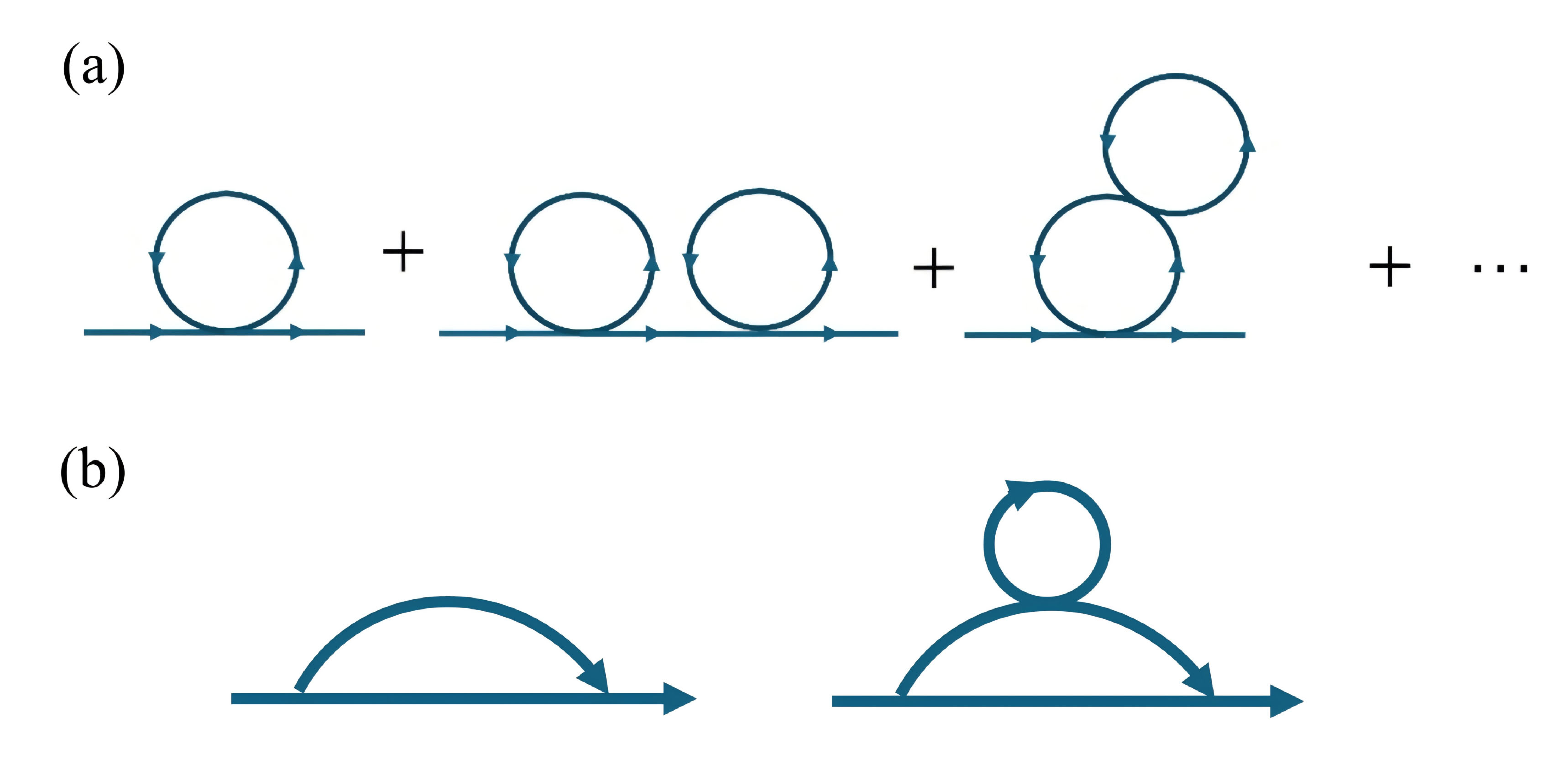}
	\caption{(a) The Feynman diagrams for the self-consistently renormalized spin wave theory. (b) The Feynman diagrams with the higher-order vertex corrections.}\label{fig6}
\end{figure}

In the above theory, we only considered the four-body magnon interaction terms and neglected others. We apply the mean-field theory to deal with MMIs and employ self-consistently renormalized spin wave theory to discuss the corrections on the magnon bands and topology in a wider temperature range. The Feynman diagram for the self-consistent theory is shown in Fig.~\ref{fig6} (a). In other models, three magnon interaction terms may arise (absent in our model), contributing other Feynman diagrams [Fig.~\ref{fig6} (b)]. When the temperature is very close to the Curi\'{e} temperature, higher-order vertex corrections and other diagrams become increasingly relevant and contribute other corrections.  

Although we only considered the limited diagrams, we believe our results capture the main physics since the critical temperatures are not that close to the Curi\'{e} temperature. The higher-order vertex corrections would make some corrections to the results, for example, the critical temperatures of topological phase transitions. The conclusions on the topological phase transitions of magnons at finite temperatures are still valid when considering MMIs.

\section{summary}
In summary, by extending the previous model~\cite{KimHongseok} and discussing the effects of MMIs on the topological phase of magnons, we demonstrated that increasing temperature can induce topological magnons at finite temperatures in honeycomb ferromagnets while the linear spin wave theory predicts a trivial phase. These finite temperature topological magnons are realized by introducing
sublattice asymmetric terms such as easy-axis anisotropy and sublattice magnetization. We apply the self-consistently renormalized spin wave theory
to deal with the MMIs, which gives rise to self-energy corrections to magnons at the mean-field level.
The emergence of the chiral edge states could be the signatures of the topological transitions, while the thermal Hall effect of magnons
can not be the signatures. Our work provides a realistic scheme for discussing the interplay between the magnon topology and MMIs.
The findings generate new insights on the potential applications of the topological magnons in magnonics.

\begin{acknowledgments}
This work is supported by the NSFC under Grant No. 12474050,  the MOST of China under Grant No. 2022YFA1204700
and Xiamen University with the startup funding.
\end{acknowledgments}

\bibliographystyle{plain}	
\bibliography{bibliography.bib}

\end{document}